\documentclass[11pt,a4paper]{article}
\usepackage{amsfonts}
\usepackage{amsmath}
\usepackage{amsthm}
\usepackage{amssymb}
\usepackage{graphicx}
\usepackage{epstopdf}

\newif\iffigs\figstrue
\begin{document}

\begin{titlepage}
\vspace{.3cm} \vspace{1cm}
\begin{center}
{\Large \textsc{\ \\[0pt]
\vspace{5mm}Quantum Fluctuations and New Instantons II: \\
Quartic Unbounded Potential} }\\[0pt]


\vspace{35pt} \textsc{V. F. Mukhanov$^{~a,b}$, E. Rabinovici$^{~c}$ and A.
S. Sorin$^{~d,e,f}$}\\[15pt]

{${}^{a}$ Ludwig Maxmillian University, \\[0pt]
Theresienstr. 37, 80333 Munich, Germany\\[0pt]
}e-mail: {\small \textit{mukhanov@physik.lmu.de}}\vspace{10pt}

{$^{b}${\small Korea Institute for Advanced Study\\[0pt]
Seoul, 02455, Korea}}\vspace{10pt}

{${}^{c}${\small Racah Institute of Physics, \\[0pt]
The Hebrew University of Jerusalem, 91904, Israel\\[0pt]
}}e-mail: {\small \textit{eliezer@vms.huji.ac.il}}\vspace{10pt}

{${}^{d}${\small Bogoliubov Laboratory of Theoretical Physics\\[0pt]
Joint Institute for Nuclear Research \\[0pt]
141980 Dubna, Moscow Region, Russia \\[0pt]
}}e-mail: {\small \textit{sorin@theor.jinr.ru}}\vspace{10pt}

{${}^{e}${\small National Research Nuclear University MEPhI\\[0pt]
(Moscow Engineering Physics Institute),\\[0pt]
Kashirskoe Shosse 31, 115409 Moscow, Russia}}\vspace{10pt}

{${}^{f}${\small Dubna State University, \\[0pt]
141980 Dubna (Moscow region), Russia}}\vspace{10pt}

\vspace{3mm}
\end{center}

\vspace{2cm}

\begin{center}
\bf{Abstract}
\end{center}
{
We study the fate of a false vacuum in the case of a potential that contains a portion which is quartic and unbounded. We first prove that an $O(4)$ invariant instanton with the Coleman boundary conditions does not exist in this case. This, however, does not imply that the false vacuum does not decay. We show how the quantum fluctuations may regularize the singular classical solutions. This gives rise to a new class of $O(4)$ invariant regularized instantons which describe the vacuum instability in the absence of the Coleman instanton. We derive the corresponding solutions and calculate the decay rate they induce.
}
\end{titlepage}

\section{Introduction}

A false vacuum, corresponding to a local minimum of the scalar field
potential $V\left( \varphi \right) ,$ is unstable and decays via tunneling.
The main contribution to the decay rate is suggested to be given by the
symmetric $O\left( 4\right) -$ solutions for the scalar field in Euclidean
time, called instantons, which satisfy the equation \cite{Coleman}:
\begin{equation}
\ddot{\varphi}(\varrho )+\frac{3}{\varrho }\,\dot{\varphi}(\varrho )-\frac{dV%
}{d\varphi }\,=0\,,  \label{1a}
\end{equation}%
where a dot denotes the derivative with respect to $\varrho =\sqrt{\tau ^{2}+%
\mathbf{x}^{2}}$, $\tau =it$ is the Euclidean time and $\mathbf{x=}\left(
x^{1},x^{2},x^{3}\right) $ are the spatial coordinates. At the instant $\tau
=0,$ the instanton emerges from under the barrier as a bubble of size $%
\varrho _{0}$ and expands. The probability of the decay per unit time and
unit volume can then be estimated as%
\begin{equation}
\Gamma \simeq \varrho _{0}^{-4}\exp \left( -S_{I}\right) \,,  \label{2a}
\end{equation}%
where%
\begin{equation}
S_{I}\,=\,2\,\pi ^{2}\,\int_{0}^{+\infty }d\varrho \,\varrho ^{3}\,\left(
\frac{1}{2}\,\dot{\varphi}^{2}\,+\,V(\varphi )\right)  \label{3a}
\end{equation}%
is the instanton action (for details see \cite{Coleman} and, for example,
\cite{Weinberg}, \cite{Mukhanov1}, \cite{MRS1}). Equation $\left( \ref{1a}%
\right) $ is an ordinary second order differential equation and requires two
boundary conditions, which according to \cite{Coleman} must be taken as%
\begin{equation}
\varphi \left( \varrho \rightarrow \infty \right) =\varphi _{0}\,,
\label{4a}
\end{equation}%
where $\varphi _{0}$ is the local false minimum of $V(\varphi ),$ and
\begin{equation}
\dot{\varphi}(\varrho =0)=0\,.  \label{5a}
\end{equation}%
The second condition is imposed to avoid a singularity in the
\textquotedblleft center of the bubble\textquotedblright\ and to assure a
finite instanton action. The condition $\left( \ref{5a}\right) ,$ formulated
in the deep ultraviolet region $\left( \varrho \rightarrow 0\right) ,$
reduces the infinite number of $O\left( 4\right)$ invariant solutions of
equation $\left( \ref{1a}\right) $ to a single non-singular solution. This,
however presents challenges when describing the false vacuum decay for a
broad class of potentials.

In particular, in $\cite{MRS1}$ we have demonstrated that for the case of a
potential with a false vacuum and a steep unbounded linear portion, the
usage of the Coleman boundary conditions ad litera leads to questionable
results. For example, in the case of a very steep unbounded potential the
Coleman instanton leads to unexpectedly high decay rate via small size
instantons It was shown that this problem is resolved once one recognizes
the inevitable effects of the quantum fluctuations, which impose both the
ultraviolet and the infrared cutoffs on the range of the validity of
classical solution. In particular, the presence of the ultraviolet cutoff,
determined by the parameters of the classical solution, allows us to abandon
the need to implement condition (\ref{5a}) as the regularization has removed
the potential singularity. As a result we obtain a new class of $O(4)$
invariant regularized solutions each of which contributes to the decay rate.

In this paper we apply these considerations to a class of potentials that
contain both a false vacuum and a portion of an unbounded quartic potential.
This potential is nearer in form to that in the standard model. First we
prove that the instantons with the Coleman boundary conditions do not exist
in this case, which in turn could lead to the erroneous conclusion that the
false vacuum is either stable which looks rather unlikely or it does not
decay via $O\left( 4\right) $ instantons. We then proceed to show how this
is resolved by taking into account the non-singular regularized instantons
and we end up by calculating the contribution of these new instantons to the
vacuum decay rate.

\section{Model}

\begin{figure}[hbt]
\begin{center}
\includegraphics[height=60mm]{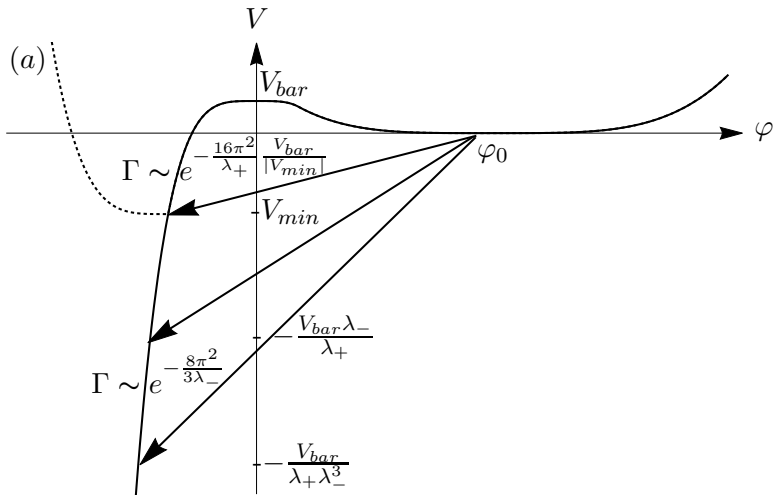} 
\end{center}
\par
\vspace*{-4.55cm} \hspace*{6.25cm} -$V_{min}$%
\par
\vspace*{-2.49cm} \hspace*{3.065cm} $(a)$%
\par
\vspace*{-1.0cm} \hspace*{6.3cm}$V$%
\par
\vspace*{1.23cm} \hspace*{9.35cm}$\varphi_{0}$%
\par
\vspace*{-1.35cm} \hspace*{6.48cm}$V_{bar}$%
\par
\vspace*{0.1cm} \hspace*{13.0cm}$\varphi$%
\par
\vspace*{2.3cm} \hspace*{6.4cm}- $-\frac{V_{bar}\lambda_-}{\lambda_+}$%
\par
\vspace*{1.08cm} \hspace*{6.4cm}-$-\frac{V_{bar}}{\lambda_+ \lambda_-^3}$%
\par
\vspace*{-4.75cm} \hspace*{4.55cm} $\Gamma\thicksim e^{-\frac{16 \pi^2}{%
\lambda_+}\frac{V_{bar}}{|V_{min}|}}$%
\par
\vspace*{2.2cm} \hspace*{4.15cm} $\Gamma\thicksim e^{-\frac{8 \pi^2}{3
\lambda_-}}$%
\par
\vspace*{1.5cm}  
\caption{\textit{A potential that contains a false vacuum and
a quartic unbounded portion is displayed for the case
 $\protect\lambda _{+}\ll \protect\lambda _{-}$. The values indicated on the lower part of vertical axis are
the values of the potential at the core of the instanton. The arrows
point to the different values of this potential associated with different values of
the parameter $E_-$. The decay probabilities $%
\Gamma$ correspond to the results obtained in equation (%
\protect\ref{48c}).} }
\label{Figure1}
\end{figure}

Let us consider the unbounded potential shown in Fig. \ref{Figure1}

\begin{equation}
V\left( \varphi \right) =\left\{
\begin{array}{ccc}
\frac{\lambda_{+}}{4}\left( \varphi -\varphi_{0}\right) ^{4} & \text{for}
& \varphi >\beta\, \varphi _{0} \\
-\frac{\lambda_{-}}{4}\left( \varphi ^{4}-\beta^{3} \varphi _{0}^{4}\right)
& \text{for} & \varphi <\beta \,\varphi _{0}\,,%
\end{array}%
\right.  \label{q1}
\end{equation}%
where%
\begin{equation}
\beta =\frac{\lambda _{+}^{1/3}}{\lambda _{+}^{1/3}+\lambda _{-}^{1/3}}\,,
\label{q2}
\end{equation}%
we consider the case when both $\lambda _{+}$ and $\lambda _{-}$ are much
smaller than unity to neglect one-loop quantum contribution to the
potential. This potential has a continuous first derivative at the point $%
\varphi _{m}=\beta \varphi _{0}$. A false vacuum local minimum is located at
$\varphi _{0}$, the potential is positive in the range $0<\varphi <\varphi
_{0},$ reaching its maximum value, the height of the barrier,
\begin{equation}
V_{bar}=\frac{\lambda _{-}\,\beta ^{3}\,\varphi_{0}^{4}}{4}  \label{q3}
\end{equation}%
at $\varphi =0$ and is unbounded for $\varphi <0.$ Let us notice that the
potential $\left( \ref{q1}\right) $ does not satisfy the conditions of the
theorem for the necessary existence of the Coleman instanton, as proven in
\cite{Coleman1}.

\vspace*{0.8cm}

\section{Absence of the Coleman Instanton}

The false vacuum corresponding to the local minimum at $\varphi _{0}$ in
Fig. \ref{Figure1} must be quantum mechanically unstable. The tunneling in
the case of the inverted quartic potential is a rather challenging problem
(see, for example, \cite{Espinosa} and references therein). First, let us prove
that the instanton with the Coleman boundary conditions does not exist for
the potential $\left( \ref{q1}\right) $ and hence cannot be responsible for
the false vacuum instability.

For $\varphi >\varphi _{m}=\beta \,\varphi _{0}$ the scalar field equation $%
\left( \ref{1a}\right) $ becomes
\begin{equation}
\ddot{\varphi}+\frac{3}{\varrho }\,\dot{\varphi}-\lambda _{+}\left( \varphi
-\varphi_{0}\right) ^{3}=0\,,  \label{q4a}
\end{equation}%
and its solution satisfying the boundary condition $\left( \ref{4a}\right) $
is
\begin{equation}
\varphi \left( \varrho \right) =\varphi_{0}\,\frac{\varrho ^{2}-\varrho
_{0}^{2}}{\varrho ^{2}-\varrho _{0}^{2}/(1+\delta )}\,,\text{ \ \ }\delta =%
\frac{4\left( 1+\sqrt{1+\lambda_{+}\,\varphi_{0}^{2}\,\varrho _{0}^{2}/2}%
\right) }{\lambda_{+}\,\varphi_{0}^{2}\,\varrho _{0}^{2}}\,,  \label{q4b}
\end{equation}%
where $\varrho _{0}^{2}$ is the remaining constant of integration yet to be
fixed. This solution must be smoothly matched to a solution of the equation
valid for $\varphi<\varphi_m$
\begin{equation}
\ddot{\varphi}+\frac{3}{\varrho }\,\dot{\varphi}+\lambda _{-}\,\varphi
^{3}=0\,,  \label{q4}
\end{equation}%
at $\varphi =\varphi _{m}.$ For the Coleman instanton the solution of $\left( %
\ref{q4}\right) $ must satisfy the boundary condition $\left( \ref{5a}%
\right) .$ Let us show that the required solution, which matches $\left( \ref%
{q4b}\right) $ and simultaneously satisfies $\left( \ref{5a}\right) $ does
not exists. With this purpose we first rewrite equation $\left( \ref{q4}%
\right) $ using new variables,
\begin{equation}
\phi =\varphi\, \varrho \,\,,\quad \eta =\ln \varrho\, ,  \label{q5}
\end{equation}%
in terms of which this equation becomes
\begin{equation}
\phi ^{\prime \prime }-\phi +\lambda _{-}\,\phi ^{3}=0\,,  \label{q6}
\end{equation}%
where a prime denotes the derivative with respect to $\eta .$ The first
integral of this equation is
\begin{equation}
E_{-}=(\phi ^{\prime })^{2}-\phi ^{2}+\frac{\lambda _{-}}{2}\,\phi ^{4}\,,
\label{q7}
\end{equation}%
which once rewritten in terms of the original variables is
\begin{equation}
E_{-}=\varrho ^{4}\,\dot{\varphi}^{2}+2\,\varrho ^{3}\,\varphi \,\dot{\varphi}+%
\frac{\lambda _{-}}{2}\,\varrho ^{4}\,\varphi ^{4}\,.  \label{q8}
\end{equation}%
According to $\left( \ref{5a}\right) $ $\dot{\varphi}$ at $\varrho =0$ must
vanish for the Coleman instanton and hence $E_{-}=0.$ On the other hand it
is required that the derivative of the solution $\left( \ref{q4b}\right)
$ which is positive at $\varphi =\varphi _{m}>0$, must be continuously
matched to the derivative of the solution of equation $\left( \ref{q4}%
\right) $ at some finite $\varrho _{m},$ where $\varphi \left( \varrho
_{m}\right) =\varphi _{m}.$ This implies that the expression in the right
hand side of $\left( \ref{q8}\right) $ must be positive, that is, $E_{-}>0$
in contradiction with $E_{-}=0$ imposed by $\left( \ref{5a}\right)$. That
completes the proof of non-existence of instanton with the Coleman boundary
conditions for the unbounded quartic potential, shown in Fig. \ref{Figure1}.

\section{Quantum Fluctuations}

As we argued in \cite{MRS1} the classical instanton can be trusted only in
that region where the contribution of the classical field and its derivative
exceed the level of minimal quantum fluctuations in the corresponding
scales, which can be roughly estimated to be (see, for example, \cite%
{Mukhanov2}):
\begin{equation}
\left\vert \delta \varphi _{q}\right\vert \simeq \frac{\sigma }{\varrho }\,,%
\text{ }\left\vert \delta \dot{\varphi}_{q}\right\vert \simeq \frac{\sigma }{%
\varrho ^{2}}\,,  \label{q8a}
\end{equation}%
where $\sigma $ is the numerical coefficient of order unity (we use $\hbar=1$ units). For quantum fluctuations $\varrho ^{4}\,\delta \dot{\varphi}%
_{q}^{2}\simeq \sigma ^{2}$ and $\left\vert \varrho ^{3}\delta \dot{\varphi}%
_{q}\delta \varphi _{q}\right\vert \simeq \sigma ^{2},$ and, therefore, as
it follows from $\left( \ref{q8}\right) ,$ for $\varphi <\varphi _{m}$ the
classical instanton solution makes sense only if
\begin{equation}
E_{-}>3\,\sigma ^{2}.  \label{q9}
\end{equation}%
The corresponding solution with positive $E_{-}$ is given in terms of Jacobi
elliptic functions and diverges as $\varrho \rightarrow 0.$ However, as we
have shown in \cite{MRS1}, this solution $\varphi \left( \varrho \right) $
is valid only at $\varrho >$ $\varrho _{uv}$ where $\varrho _{uv}$ is the
ultraviolet cutoff scale determined by the condition%
\begin{equation}
\dot{\varphi}\left( \varrho _{uv}\right) \simeq \frac{\sigma }{\varrho
_{uv}^{2}}\,.  \label{q9a}
\end{equation}%
The solution of this equation for ultraviolet cutoff is presented in the next
section. This cutoff regularizes the instantons which are singular otherwise
and have infinite action$.$ 

\section{New Instantons}

The calculations leading to the explicit form of the $O(4)$ invariant
regularized solution are somewhat more involved in the quartic case compared
to those in the unbouded linear behavior case \cite{MRS1}. We now sketch the various steps involved.
First one identifies the parameter on which the solution depends. One then
proceeds to find the ultraviolet cutoff scale $\varrho _{uv}$ (the inftrared
cutoff will not play an important role) in terms of the parameters of the
solution. Next we will constructs the explicit regularized solutions valid
in the range above the ultraviolet cutoff scale. Equipped with these
solutions one determines the action and the false vacuum decay rate. This
involves various Jacobi elliptic functions. Finally we will present the
results for various asymptotic limits.

The ultraviolet cutoff scale regularizes a whole class of new instantons
which can be parametri-zed either by $E_{-}$ or alternatively by $\varrho
_{0}\left( E_{-}\right) .$ It is convenient to parametrize them by a new
variable $\chi $ related to $\varrho _{0}$ in $\left( \ref{q4b}\right) $ as%
\begin{equation}
\varrho _{0}^{2}\,\varphi _{0}^{2}=\frac{E_{c}}{4\left( 1-\beta \right) }%
\,\chi \,\left( 1-\beta +\chi \right) \,,  \label{q11}
\end{equation}%
where%
\begin{equation}
E_{c}\equiv \frac{32}{\lambda _{+}\,(1-\beta )}\,.  \label{q11a}
\end{equation}%
Substituting $\left( \ref{q11}\right) $ in the expression for $\delta $ in $%
\left( \ref{q4b}\right) $ we find%
\begin{equation}
\delta =\frac{ 1-\beta }{\chi }\,.  \label{q12}
\end{equation}%
The solution $\left( \ref{q4b}\right) $ is valid only $\varrho >\varrho
_{m}, $ where $\varrho _{m}$ is determined by
\begin{equation}
\varphi \left( \varrho _{m}\right) \equiv \varphi _{m}=\beta \,\varphi
_{0}\,,  \label{q12a}
\end{equation}%
from where one gets%
\begin{equation}
\varrho _{m}^{2}\,\varphi _{0}^{2}=\frac{E_{c}}{4\left( 1-\beta \right) }\,\chi\,
\left( 1+\chi \right) \,.  \label{q13}
\end{equation}%
At $\varrho =\varrho _{m}$ we have to match $\left( \ref{q4b}\right) $ to a
solution of equation $\left( \ref{q4}\right) $ with $E_{-}>0.$ This solution
is given in terms of the Jacobi elliptic functions \cite{BE},%
\begin{equation}
\varphi (\varrho )=\frac{\sqrt{2}\,k}{\varrho \,\sqrt{\lambda
_{-}\,(2\,k^{2}-1)}}\,\text{cn}\left( \frac{\ln \left( \varrho /\alpha
\right) }{\sqrt{2\,k^{2}-1}},k\right)\, ,  \label{q14}
\end{equation}%
where
\begin{equation}
k\equiv \frac{1}{\sqrt{2}}\,\sqrt{1+\frac{1}{\sqrt{1+2\,\lambda _{-}\,E_{-}}}%
}\,,\quad \frac{1}{\sqrt{2}}<k\leq 1\,,  \label{q15}
\end{equation}%
and the constants of integration $\alpha $ and $E_{-}$ are determined in
terms of $\varrho _{0}^{2}$ or, alternatively $\chi ,$ by requiring that the
field $\varphi $ and it's first derivative are continuous at $\varrho_{m}.$

First let us find $E_{-}\left( \chi \right) .$ The expression $\left( \ref%
{q8}\right) $ for $E_{-}$ rewritten in terms of the field values at the
matching point $\varrho _{m}$ becomes%
\begin{equation}
E_{-}=\varrho _{m}^{4}\,\dot{\varphi}_{m}^{2}+2\,\varrho _{m}^{3}\,\dot{%
\varphi}_{m}\varphi _{m}+\frac{\lambda _{-}}{2}\,\varrho _{m}^{4}\,\varphi
_{m}^{4}\,.  \label{q16}
\end{equation}%
On the other hand, the first integral of equation $\left( \ref{q4a}\right) $
expressed in terms of the field values at the same point is equal%
\begin{equation}
E_{+}=\varrho _{m}^{4}\,\dot{\varphi}_{m}^{2}+2\,\varrho _{m}^{3}\,\left( \varphi _{m}-\varphi _{0}\right)\,\dot{%
\varphi}_{m} -\frac{\lambda _{+}}{2}%
\,\varrho _{m}^{4}\,\left( \varphi _{m}-\varphi _{0}\right) ^{4}\,,
\label{q17}
\end{equation}%
and it vanishes for the solution $\left( \ref{q4b}\right) $. Keeping this in
mind and subtracting $\left( \ref{q17}\right) $ from $\left( \ref{q16}%
\right) $ we find%
\begin{equation}
E_{-}\left( \chi \right) =E_{c}\,\chi \,\left( 1+\chi \right) ^{3} \,.
\label{q18}
\end{equation}
   From here it follows that the quantum bound $\left( \ref{q9}\right) $
imposes the following lower bound on the possible values of $\chi $%
\begin{equation}
\chi >\chi _{c}\simeq \frac{3\,\sigma ^{2}}{E_{c}} \,.  \label{q18d}
\end{equation}%
Because the field must be continuous at $\rho _{m}$ the constant of
integration $\alpha $ is determined by solving the equation we obtain after
substitution $\left( \ref{q14}\right) $ in $\left( \ref{q12a}\right) $:
\begin{equation}
\text{cn}\left( \frac{\ln \left( \varrho _{m}/\alpha \right) }{\sqrt{%
2\,k^{2}-1}},k\right) =\sqrt{\lambda_{-}\,\beta ^{2}\,\varrho _{m}^{2}\,\varphi
_{0}^{2}\,\left( 1-\frac{1}{2k^{2}}\right) }\equiv \varepsilon _{m},\ \ \text{%
sn}\left( \frac{\ln \left( \varrho _{m}/\alpha \right) }{\sqrt{2\,k^{2}-1}}%
,k\right) <0 \,,  \label{q18a}
\end{equation}
where both $\varrho _{m},$ and $k$ can be expressed through $\chi .$ The
classical solution $\left( \ref{q14}\right) $ has a bounce at $\varrho
_{b}<\varrho _{m},$ at which $\dot{\varphi}\left( \varrho _{b}\right) =0.$
Equating the derivative of $\left( \ref{q14}\right) $ to zero, one gets the
following equation for $\varrho _{b}$:

\begin{equation}
\text{cn}\left( \frac{\ln \left( \varrho _{b}/\alpha \right) }{\sqrt{%
2\,k^{2}-1}},k\right) =-\left( \frac{1}{k^{2}}-1\right) ^{\frac{1}{4}}\equiv
-\varepsilon _{b}\,,\text{ \ \ \ sn}\left( \frac{\ln \left( \varrho
_{b}/\alpha \right) }{\sqrt{2\,k^{2}-1}},k\right) >0\,.  \label{q20}
\end{equation}%
Let us notice the following useful relation between $\varepsilon _{m}$ and $%
\varepsilon _{b},$ which appear in equations $\left( \ref{q18a}\right) $ and
$\left( \ref{q20}\right) $:%
\begin{equation}
\varepsilon _{m}=f\left( \chi \right) \,\varepsilon_{b}\,,  \label{q20c}
\end{equation}%
where
\begin{equation}
f\left( \chi \right) =\left( \frac{\beta \,\chi }{1+\chi }\right) ^{1/4}<1\,.
\label{q20d}
\end{equation}

The solution $\left( \ref{q14}\right) $ fails before the bounce is reached.
In fact, at $\varrho _{uv}>$ $\varrho _{b},$ determined by $\left( \ref{q9a}%
\right) ,$ the quantum fluctuations become dominant and the bubble emerges
from under the barrier. Introducing $\varepsilon _{uv},$ defined as%
\begin{equation}
\varepsilon _{uv}\equiv -\text{cn}\left( \frac{\ln \left( \varrho
_{uv}/\alpha \right) }{\sqrt{2\,k^{2}-1}},k\right) \quad \text{for} \text{ \
\ \ sn}\left( \frac{\ln \left( \varrho _{uv}/\alpha \right) }{\sqrt{%
2\,k^{2}-1}},k\right) >0\,,  \label{q21ab}
\end{equation}%
and substituting $\left( \ref{q14}\right) $ in $\left( \ref{q9a}\right) $ we
find that $\varepsilon _{uv}$ satisfies the following equation
\begin{equation}
\left( \frac{\varepsilon _{uv}}{\varepsilon _{b}}\right) ^{4}-\left( \frac{%
32\,\sigma^{4}}{\lambda_{-}\,E_{-}^{3}}\right) ^{1/4}\frac{\varepsilon _{uv}%
}{\varepsilon _{b}}-\left( 1-\frac{\sigma ^{2}}{E_{-}}\right) =0\,,
\label{q22ac}
\end{equation}%
the exact solution of which is
\begin{equation}
\frac{\varepsilon _{uv}}{\varepsilon _{b}}=\left( 1-\frac{\sigma ^{2}}{E_{-}}%
\right)^{1/4}W\left( \sqrt{1-\frac{\gamma }{\sqrt{27}\,W^{6}}}+\sqrt{\frac{%
\gamma }{\sqrt{27}\,W^{6}}}\right) \,,  \label{q22ad}
\end{equation}%
where%
\begin{equation}
\gamma =\sqrt{\frac{27\,\sigma^{4}}{8\,\lambda _{-}E_{-}^{3}}\left( 1-\frac{%
\sigma^{2}}{E_{-}}\right) ^{-1}}\,,  \label{q22af}
\end{equation}%
and%
\begin{equation}
W=\left[ \frac{1}{3}\left( 1+\left( \sqrt{1+\gamma ^{2}}+\gamma \right)
^{2/3}+\left( \sqrt{1+\gamma ^{2}}-\gamma \right) ^{2/3}\right) \right]
^{1/4}.  \label{q22ag}
\end{equation}%
Now keeping in mind the signs of the Jacobi sn\textbf{(}$x,k$\textbf{)} in $%
\left( \ref{q18a}\right) ,\left( \ref{q20}\right) $ and $\left( \ref{q21ab}%
\right) $ we first find that the elliptic Jacobi amplitudes are%
\begin{eqnarray}
&&\text{am}\left( \frac{\ln \left( \varrho _{m}/\alpha \right) }{\sqrt{%
2k^{2}-1}},k\right) =2\,\pi -\arccos \left( \varepsilon _{m}\right) \,,\,
\notag \\
&&\text{am}\left( \frac{\ln \left( \varrho _{uv}/\alpha \right) }{\sqrt{%
2k^{2}-1}},k\right) =\arccos \left( -\varepsilon _{uv}\right) ,  \notag \\
&&\,\text{am}\left( \frac{\ln \left( \varrho _{b}/\alpha \right) }{\sqrt{%
2k^{2}-1}},k\right) =\arccos \left( -\varepsilon _{b}\right) ,
\end{eqnarray}%
and then obtain the following expressions for the ratio of the corresponding
scales%
\begin{equation}
\ln \frac{\varrho _{m}}{\varrho _{b}}=\sqrt{2\,k^{2}-1}\left( 2\,\boldsymbol{%
K}(k)+F(\pi -\arccos (-\varepsilon _{b}),k)-F(\arccos \left( \varepsilon
_{m}\right) ,k)\right)\,,  \label{q22bc}
\end{equation}%
and%
\begin{equation}
\ln \frac{\varrho _{m}}{\varrho _{uv}}=\sqrt{2\,k^{2}-1}\left( 2\,%
\boldsymbol{K}(k)+F(\pi -\arccos (-\varepsilon _{uv}),k)-F(\arccos
(\varepsilon_{m}),k)\right)\, ,  \label{q22bd}
\end{equation}%
where $F\left( x,k\right) $ and $\boldsymbol{K}(k)$ are incomplete and
complete elliptic integral of the first kind correspondingly.

Taking into account the definitions of $\varepsilon _{m}$ and $\varepsilon
_{uv}$ in $\left( \ref{q18a}\right) $ and $\left( \ref{q21ab}\right) $ we
find from $\left( \ref{q14}\right) $ the following useful expressions for
the value of the field on the border of quantum core$,$%
\begin{equation}
\varphi _{uv}\equiv \varphi \left( \varrho _{uv}\right) =-\frac{\varrho _{m}%
}{\varrho _{uv}}\frac{\varepsilon _{uv}}{\varepsilon_{m}}\,\beta \,\varphi
_{0}\,,  \label{q22abc}
\end{equation}%
and for the potential%
\begin{equation}
V_{uv}\equiv V\left( \varphi _{uv}\right) =-V_{bar}\left( \beta \left( \frac{%
\varrho _{m}}{\varrho _{uv}}\frac{\varepsilon _{uv}}{\varepsilon _{m}}%
\right) ^{4}-1\right) ,  \label{q22abcd}
\end{equation}%
correspondingly.

Finally let us calculate the action for the regularized $\chi -$ instantons.
Substituting $\left( \ref{q4b}\right) $ and $\left( \ref{q14}\right) $ into%
\begin{equation}
S_{I}\,=\,2\,\pi ^{2}\,\int_{\varrho _{uv}}^{+\infty }d\varrho \,\varrho
^{3}\,\left( \frac{1}{2}\,\dot{\varphi}^{2}\,+\,V(\varphi )\right) +\frac{%
\pi ^{2}}{2}V\left( \varphi \left( \varrho _{uv}\right) \right) \varrho
_{uv}^{4}\,,  \label{q23aa}
\end{equation}%
where the second term account for the contribution of quantum core (see,
\cite{MRS1} for justification), we obtain

\begin{eqnarray}
S_{I}\left( \chi \right)  &=&\frac{\pi ^{2}}{3}\left[ \frac{E_{-}}{4}\left(
3+\frac{1}{1+\chi }+\frac{1-2\,\beta}{\left( 1-\beta \right)
\left( 1+\chi \right)^{2}}+\frac{1-\beta}{\left( 1+\chi
\right) ^{3}}\right) \right.   \notag \\
&&+\left. \frac{4\left( 1+2\,\lambda_{-}\,E_{-}\right) ^{1/4}}{\,\lambda _{-}\,}%
\left( 4\,\boldsymbol{E}(k)-\boldsymbol{E}\left( \arccos \left( \varepsilon
_{m}\right) ,k\right) -\boldsymbol{E}\left( \arccos \left( -\varepsilon
_{uv}\right) ,k\right) \right) \right.   \notag \\
&&+\left. \frac{E_{-}\,(2-3\,k^{2})}{k^{2}}\ln \left( \frac{\varrho _{m}}{%
\varrho _{uv}}\right) +\left( \frac{2E_{-}}{\lambda _{-}}\right)
^{1/2}\left( \frac{\varepsilon _{uv}}{\varepsilon _{b}}\right) ^{2}+2\,\sigma
\left( \frac{2\,E_{-}}{\lambda _{-}}\right) ^{1/4}\frac{\varepsilon _{uv}}{%
\varepsilon _{b}}\right] ,  \label{q21}
\end{eqnarray}%
where $E_{-}$ can be expressed in terms of $\chi $ and vise versa using $%
\left( \ref{q18}\right) $, $\varepsilon _{uv}$ is the solution of equation $%
\left( \ref{q22ac}\right) $ related to $\varrho _{uv}$ as in $\left( \ref%
{q21ab}\right) ,$ $\varepsilon _{b}$ is defined in $\left( \ref{q20}\right) ,
$ and $\boldsymbol{E}\left( x,k\right) $ and $\boldsymbol{E}(k)$ are
incomplete and complete elliptic integral of the second kind, respectively.

\section{Asymptotics}

The exact solutions above are not very transparent and therefore it is
useful to consider the limiting cases for which the special Jacobi functions
simplify to elementary functions.

Let us begin with displaing the asymptotics of the exact solution $\left( %
\ref{q22ad}\right) ,$ which contains the parameter $\gamma ,$ defined in $%
\left( \ref{q22af}\right) .$ Because $\gamma \sim \left( \lambda
_{-}E_{-}^{3}\right) ^{-1/2}$, there are two cases we have to consider, namely,
$\lambda _{-}^{-1/3}\gg E_{-}>3\sigma ^{2}$ and $E_{-}\gg \lambda
_{-}^{-1/3},$ which correspond to $\gamma \gg 1$ and $\gamma \ll 1$
respectively. The asymptotics are%
\begin{eqnarray}
\frac{\varepsilon _{uv}}{\varepsilon _{b}} &\simeq &\left( \frac{32\,\sigma
^{4}}{\lambda _{-}\,E_{-}^{3}}\right) ^{1/12}\text{ \ \ \ \ \ \ \ \ for \ \
\ \ }\frac{1}{\lambda _{-}^{1/3}}\gg E_{-}>3\sigma ^{2}\,,  \notag \\
&\simeq &1+\left( \frac{\sigma ^{4}}{8\,\lambda_{-}\,E_{-}^{3}}\right)
^{1/4} \, \, \text{ \ \ for \ \ \ \ \ \ }E_{-}\gg \frac{1}{\lambda _{-}^{1/3}%
}\,.  \label{q50a}
\end{eqnarray}%
As we see in the first case $\varepsilon _{uv}$ can be significantly larger
that $\varepsilon _{b}.$ For example for the minimal possible $E_{-}\sim
3\sigma ^{2}$, allowed by quantum limit, $\varepsilon _{uv}/\varepsilon
_{b}\simeq \left( \sigma ^{2}\lambda _{-}\right) ^{-1/12}$ can be larger
than unity but only for very small $\lambda _{-}.$ For $E_{-}\gg 1/\lambda
_{-}^{1/3}$ the instant at which the classical description fails is always
very close to the moment of the bounce.

\begin{figure}[hbt]
\begin{center}
\includegraphics[height=60mm]{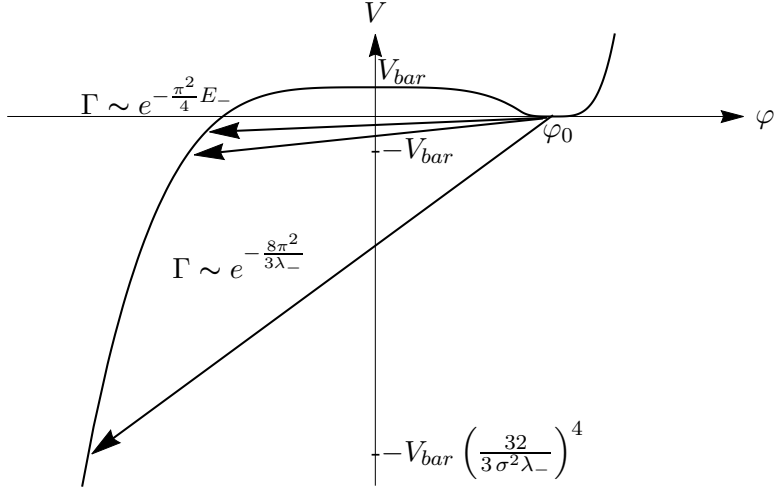} 
\end{center}
\par
\vspace*{-3.8cm} \hspace*{5.2cm} $\Gamma\thicksim e^{-\frac{8 \pi^2}{%
3\lambda_-}}$%
\par
\vspace*{-2.7cm} \hspace*{4.0cm} $\Gamma\thicksim e^{-\frac{\pi^2}{4}E_-}$%
\par
\vspace*{-1.7cm} \hspace*{7.85cm}$V$%
\par
\vspace*{1.05cm} \hspace*{10.2cm}$\varphi_{0}$%
\par
\vspace*{-1.246cm} \hspace*{8.02cm}$V_{bar}$%
\par
\vspace*{0.1cm} \hspace*{13.0cm}$\varphi$%
\par
\vspace*{-0.01cm} \hspace*{7.95cm}-$-V_{bar}$%
\par
\vspace*{3.4cm} \hspace*{7.95cm}-$-V_{bar}\left(\frac{32}{3\,\sigma^2
\lambda_-}\right)^4$%
\par
\vspace*{0.5cm}  
\caption{\textit{A potential in the case
 $\protect\lambda _{-}\ll \protect\lambda _{+}$.
The values indicated on the lower part of vertical axis are the values of
the potential at the core of the instanton. The corresponding values of the decay probability $\Gamma$ are
given in equation (\protect\ref%
{El_fig_2}).} }
\label{Figure2}
\end{figure}

There exist two qualitatively different classes of the potentials, namely,
those for which $\lambda _{+}\ll \lambda _{-}$ $\left( \beta \ll 1\right) $
(see Fig. \ref{Figure1}) with rather sharp maximum and the potentials with $%
\lambda _{-}\ll \lambda _{+}$ and $1-\beta \simeq \left( \lambda
_{-}/\lambda _{+}\right) ^{1/3}\ll 1$ (Fig. \ref{Figure2}) having rather
flat maximum. We will consider these two cases separately.

The behavior of elliptic functions and integrals in the formulae above
depends on the value of the parameter
\begin{equation}
2\,\lambda_{-}\,E_{-}=64\,\frac{\left( 1-\beta \right) ^{2}}{\beta ^{3}}%
\,\chi\, \left( 1+\chi \right) ^{3}.  \label{q40}
\end{equation}%
\textit{For} $2\,\lambda_{-}\,E_{-}\ll 1$
\begin{equation}
k\simeq 1-\frac{\lambda _{-}\,E_{-}}{4}  \label{q40a}
\end{equation}%
and both
\begin{equation}
\varepsilon _{b}\simeq \left( \frac{\lambda_{-}\,E_{-}}{2}\right)^{1/4}\,,
\label{q41a}
\end{equation}%
and $\varepsilon _{m}=f\varepsilon _{b}$, where $f$ is given in $\left( \ref%
{q20d}\right) ,$ are much smaller than unity, that is, $\varepsilon
_{m},\varepsilon _{b}\ll 1.$ For $k\rightarrow 1$ the elliptic intergrals
diverge logarithmically. To reveal the main non-analytic term in the
expansion of $F(\arccos \left( \varepsilon \right) ,k)$ for $\varepsilon \ll
1$ we first note that%
\begin{equation}
\arccos \left( \varepsilon \right) =\frac{\pi }{2}-\varepsilon +O\left(
\varepsilon ^{3}\right)  \label{41b}
\end{equation}%
and
\begin{equation}
F(\arccos \left( \varepsilon \right) ,k)=\int_{0}^{\frac{\pi }{2}%
-\varepsilon +O\left( \varepsilon ^{3}\right) }\frac{d\alpha }{\sqrt{%
1-k^{2}\sin ^{2}\alpha }}=\boldsymbol{K}\left( k\right)
-\int_{0}^{\varepsilon -O\left( \varepsilon ^{3}\right) }\frac{d\tilde{\alpha%
}}{\sqrt{1-k^{2}\cos ^{2}\tilde{\alpha}}} \,.  \label{42}
\end{equation}%
Keeping in mind that in the whole range of integration $\tilde{\alpha}$ $\ll
1$ we can expand $\cos ^{2}\tilde{\alpha}$ in series and the integral in $%
\left( \ref{42}\right) $ can be approximated as
\begin{eqnarray}
&&\int_{0}^{\varepsilon -O\left( \varepsilon ^{3}\right) }\frac{1}{\sqrt{%
1-k^{2}+k^{2}\,\tilde{\alpha}^{2}}}\left( 1+O\left( \frac{\tilde{\alpha}^{4}%
}{1-k^{2}+k^{2}\,\tilde{\alpha}^{2}}\right) \right) d\tilde{\alpha}  \notag
\\
&=&\frac{1}{k}\,\ln \frac{k\,\varepsilon +\sqrt{1-k^{2}+k^{2}\,%
\varepsilon^{2}}}{\sqrt{1-k^{2}}}+O\left( \varepsilon ^{2}\right)\,,
\label{43}
\end{eqnarray}%
and hence $\left( \ref{42}\right) $ becomes%
\begin{equation}
F(\arccos \left( \varepsilon \right) ,k)=\boldsymbol{K}\left( k\right) -%
\frac{1}{k}\ln \frac{\varepsilon +\sqrt{\varepsilon _{b}^{4}+\varepsilon ^{2}%
}}{\varepsilon _{b}^{2}}+O\left( \varepsilon ^{2}\right)\,,  \label{44}
\end{equation}%
where we took into account the definition of $\varepsilon _{b}$ in $\left( %
\ref{q20}\right) .$ The complete elliptic integral $\boldsymbol{K}\left(
k\right) $ for $k=1-\epsilon $ has the following expansion (see \cite{BE})%
\begin{equation}
\boldsymbol{K}\left( k\right) =\frac{1}{2}\ln \frac{8}{\epsilon }+O\left(
\epsilon \ln \epsilon \right)\,.  \label{45}
\end{equation}%
\textit{For} $2\lambda _{-}E_{-}\gg 1$%
\begin{equation}
k\simeq \frac{1}{\sqrt{2}}\left( 1+\frac{1}{\sqrt{8\,\lambda _{-}\,E_{-}}}%
\right) ,\text{ \ \ }\varepsilon _{b}\simeq 1-\frac{1}{\sqrt{8\,\lambda
_{-}E_{-}}}\,.\text{\ \ }  \label{45a}
\end{equation}%
The point $k=1/\sqrt{2}$ is not singular and for example
\begin{equation}
\boldsymbol{K}\left( \frac{1}{\sqrt{2}}+\epsilon \right) =\frac{1}{4\sqrt{%
\pi }}\left[ \Gamma \left( \frac{1}{4}\right) \right] ^{2}+O\left( \epsilon
\right) \simeq 1.85+O\left( \epsilon \right) \,.  \label{45b}
\end{equation}%
Using these expansions we can simplify the exact formulae derived in the
previous section and obtain the expressions for $V_{uv},$ $\varrho _{0}^{2}$
and the action $S\,\ $\ as a function of $E_{-}$ parametrizing the
instantons. In the derivation one has to use $E_{-}$ and $\chi $
interchangeably resolving $\left( \ref{q40}\right) $ in the approximations $%
\chi \gg 1$ and $\chi \ll 1.$ Skipping details of the calculations we will
present below the results in the leading order.

$\boldsymbol{The}$ $\boldsymbol{potential}$ $\boldsymbol{with}$ $\boldsymbol{%
a}$ $\boldsymbol{sharp}$ $\boldsymbol{maximum.}$\textit{\ }First let us
consider the potential shown in Fig. \ref{Figure1} for which $\lambda
_{-}\gg \lambda _{+} $ and, hence, $\beta \ll 1.$ As it follows from $\left( %
\ref{q40}\right) $ in this case%
\begin{eqnarray}
\chi &\simeq &\frac{\lambda _{+}\,E_{-}}{32}\text{ \ \ \ \ \ \ \ \ \ \ for \
\ }\frac{32}{\lambda _{+}}\gg E_{-}>3\sigma ^{2},  \notag \\
&\simeq &\left( \frac{\lambda _{+}\,E_{-}}{32}\right) ^{1/4}\text{ \ \ for \
\ }E_{-}\gg \frac{32}{\lambda _{+}}\,  \label{47c}
\end{eqnarray}%
and substituting this in $\left( \ref{q11}\right) $ we find
\begin{eqnarray}
\varphi _{0}^{2}\,\varrho _{0}^{2} &\simeq &\frac{E_{-}}{4} \,\,\,\,\,\,%
\text{ \ \ \ for \ }\frac{32}{\lambda _{+}}\gg E_{-}>3\sigma ^{2},  \notag \\
&\simeq &\sqrt{\frac{2\,E_{-}}{\lambda_{+}}} \text{\ \ \ for \ \ }E_{-}\gg
\frac{32}{\lambda _{+}}\,.  \label{48a}
\end{eqnarray}%
The calculation of $V_{uv}$ is more involved. We have first to expand $%
\left( \ref{q22bc}\right) $ and $\left( \ref{q22bd}\right) $, using
approximations derived above, and then with taking into account $\left( \ref%
{q20c}\right) $ and $\left( \ref{q50a}\right) $ from $\left( \ref{q22abcd}%
\right) $ one gets%
\begin{eqnarray}
V_{uv} &\simeq &-V_{bar}\,\frac{\lambda _{-}}{\lambda _{+}}\,\left( \frac{32%
}{\lambda_{-}\,E_{-}}\right) ^{4} \,\,\,\,\text{ \ \ for \ }\frac{1}{%
2\lambda _{-}}\gg E_{-}>3\sigma ^{2}\,,  \notag \\
&\simeq &-V_{bar}\,\left( \frac{32}{\lambda _{+}\,E_{-}}\right) \text{ \ \ \
\ \ \ \ \ \ \ for \ }\frac{32}{\lambda _{+}}\gg E_{-}\gg \frac{1}{2\lambda
_{-}},  \notag \\
&\simeq &-V_{bar}\left( \frac{32}{\lambda_{+}\,E_{-}}\right) ^{1/4} \text{ \
\ \ \ \ \ \ for \ }E_{-}\gg \frac{32}{\lambda _{+}}\,.  \label{48b}
\end{eqnarray}%
Finally, one can check that the leading term in the action in different
asymptotics is%
\begin{eqnarray}
S &\simeq &\frac{8\,\pi ^{2}}{3\,\lambda _{-}} \text{ \ \ \ \ for \ }\frac{1%
}{2\lambda _{-}}\gg E_{-}>3\sigma ^{2},  \notag \\
&\simeq &\frac{\pi ^{2}}{2}\,E_{-}\,\, \text{ \ for \ }\frac{32}{\lambda _{+}%
}\gg E_{-}\gg \frac{1}{2\lambda _{-}}\,,  \notag \\
&\simeq &\frac{\pi ^{2}}{4}\text{\ }E_{-} \, \text{ \ for \ }E_{-}\gg \frac{%
32}{\lambda _{+}}\,.  \label{48c}
\end{eqnarray}

$\boldsymbol{The}$ $\boldsymbol{potential}$ $\boldsymbol{with}$ $\boldsymbol{%
a}$ $\boldsymbol{rather\, flat}$ $\boldsymbol{barrier.}$\textit{\ }In the
case of a flat barrier (Fig. \ref{Figure2}) with $\lambda _{-}\ll \lambda
_{+}$ and $1-\beta \ll 1$ the calculations are very similar. The solution of
$\left( \ref{q40}\right) $ is%
\begin{eqnarray}
\chi &\simeq &\frac{\lambda _{-}E_{-}}{32\left( 1-\beta \right) ^{2}}\,\,%
\text{ \ \ \ \ \ \ \ \ \ \ for \ \ }\frac{32\left( 1-\beta \right) ^{2}}{%
\lambda _{-}}\gg E_{-}>3\sigma ^{2}\,,  \notag \\
&\simeq &\left( \frac{\lambda _{-}E_{-}}{32\left( 1-\beta \right) ^{2}}%
\right) ^{1/4}\,\, \text{ \ \ for \ \ }E_{-}\gg \frac{32\left( 1-\beta
\right) ^{2}}{\lambda _{-}}\,.
\end{eqnarray}%
Correspondingly, for the size of the bubble we obtain%
\begin{eqnarray}
\varphi _{0}^{2}\varrho _{0}^{2} &\simeq &\frac{E_{-}}{4}\left( 1+\frac{%
\lambda _{+}E_{-}}{32}\right) \,\,\,\, \text{ \ \ \ for \ }\frac{32\left(
1-\beta \right) ^{2}}{\lambda _{-}}\gg E_{-}>3\sigma ^{2},  \notag \\
&\simeq &\sqrt{\frac{2E_{-}}{\lambda _{-}}}\,\, \text{ \ \ \ \ \ \ \ \ \ \ \
\ \ \ \ \ \ \ \ for \ \ }E_{-}\gg \frac{32\left( 1-\beta \right) ^{2}}{%
\lambda _{-}}\,,  \label{q52a}
\end{eqnarray}%
the potential $V_{uv}$ is given by%
\begin{eqnarray}
V_{uv} &\simeq &-V_{bar}\left( \frac{32}{\lambda _{-}E_{-}}\right) ^{4}
\text{ \ \ \ \ for \ }\frac{1}{2\lambda _{-}}\gg E_{-}>3\sigma ^{2}\,, \text{%
\ }  \notag \\
&\simeq &-V_{bar}\frac{12.45}{\left( \lambda _{-}E_{-}\right) ^{1/4}} \,\,%
\text{ \ \ \ \ for \ \ }E_{-}\gg \frac{1}{2\lambda _{-}}  \label{q52b}
\end{eqnarray}%
and the action is
\begin{eqnarray}
S &\simeq &\frac{8\,\pi^{2}}{3\,\lambda _{-}} \text{ \ \ \ \ for \ }\frac{1}{%
2\lambda _{-}}\gg E_{-}>3\sigma ^{2}\,,  \notag \\
&\simeq &\frac{\pi ^{2}}{4}\text{\ }E_{-} \text{ \ for \ }E_{-}\gg \frac{1}{%
2\,\lambda _{-}}\,.
\label{El_fig_2}
\end{eqnarray}

In both cases the contribution of the $E_{-}-$instantons to the overall
decay rate per unit volume per unit time is given by%
\begin{equation}
\Gamma \left( E_{-}\right) \sim \left(\varrho _{0}\left( E_{-}\right)\right)^{-4}\,
e^{-S\left( E_{-}\right)} \,,  \label{q45}
\end{equation}%
and from the formulae above one finds that for the unbounded potentials the
main contribution comes from instantons with $1/2\lambda
_{-}>E_{-}>3\,\sigma ^{2}$ for which the leading term in the action is $\frac{8\,
\pi^{2}}{3\,\lambda _{-}}.$ The contribution of these instantons only weakly depends
on the their size, which is always larger than $\varphi _{0}^{-1}.$

The results above can also be used to estimate the leading contribution to
the decay rate for the potentials with two minima (see for example curve (a)
in Fig. \ref{Figure1}). If the true minimum has the depth $V_{\min },$ the
corresponding $E_{-}$ which gives the largest contribution to the decay rate
is determined by the equation $V_{\min }\simeq V_{uv}\left( E_{-}\right) .$
For example, in the case $\lambda _{+}\ll \lambda _{-}$ for the true minimum
in the range%
\begin{equation}
64\,\frac{\lambda _{-}}{\lambda _{+}}\,V_{bar}\gg \left\vert V_{\min
}\right\vert \gg V_{bar}
\end{equation}%
from equations $\left( \ref{48a}\right) $ and $\left( \ref{48b}\right) $ we
get%
\begin{equation}
\Gamma \sim \left( \frac{\lambda _{+}}{8}\frac{\left\vert V_{\min
}\right\vert }{V_{bar}}\right) ^{2}\varphi _{0}^{4}\,\exp \left( -\frac{%
16\pi ^{2}}{\lambda _{+}}\frac{V_{bar}}{\left\vert V_{\min }\right\vert }%
\right) \,.
\end{equation}%
In the other cases the calculations are very similar.

\section{Conclusions}

In this paper we have considered the false vacuum decay for the unbounded
quartic potential. Using exact solutions we have proven that the instanton
with the Coleman boundary conditions does not exist in this case. One has to
point out that this result does not contradict to the theorem in \cite%
{Coleman1} because not all necessary conditions for the existence of solution
with the Coleman boundary conditions are satisfied for quartic potentials. We
suspect (although not proven this) that the Coleman instanton also does not
exist for unbounded potential $-\varphi ^{n}$ with $n>4.$ Taken literally
this result, one could have erroneously concluded that the decay of the false
vacuum cannot be described anymore by the most symmetric $O\left( 4\right) $
Euclidean solutions for these potentials.

In the previous paper \cite{MRS1} we have shown that one of the boundary
conditions used by Coleman to avoid the singularity of the classical
solution must be abandoned because it is formulated in the deep ultraviolet
region where quantum fluctuations definitely dominate and the classical
solution cannot be trusted anymore. The quantum fluctuations naturally
regularize the singular classical solutions by introducing the cutoff scale
entirely determined by the parameters of the classical solution. In turn, this
leads to the appearance of a new class of instantons, which being singular
in the absence of cutoff, were not contributing to the false vacuum decay
before. The new $O\left( 4\right) -$instantons allow us to calculate the
leading contribution to vacuum decay rate in those case when the $O\left(
4\right) -$instantons with the Coleman boundary conditions do not exist. For the
particular example of quartric unbounded potential we have found the
explicit solutions, which can also be applied to a broad class of potentials
with false and true minima separated by quartic potential.

\bigskip

\textbf{Acknowledgments}

\bigskip

The work of V. M. was supported under Germany's Excellence
Strategy---EXC-2111---Grant No. 39081486.

The work of E. R. is partially supported by the Israeli Science Foundation
Center of Excellence. E. R. would also like to thank NHETC at Rutgers Physics
Department, the IHES and the CCPP at NYU for hospitality.

A. S. would like express a gratitude to the Racah Institute of Physics of
the Hebrew University of Jerusalem and Ludwig Maxmillian University of
Munich for the hospitality during his visits. The work of A. S. was
partially supported by RFBR grant No. 20-02-00411.

\end{document}